\documentstyle[aps,epsfig,prl]{revtex}
\begin{document}
\title{Analogue of the Kubo Formula  for Conductivity of
Spatially Inhomogeneous Systems and Electric Fields.}
\author{S. T. Pavlov\dag\ddag}
\address {\dag Facultad de Fisica de la UAZ, Apartado Postal C-580,
98060 Zacatecas, Zac., Mexico \\
\ddag P.N. Lebedev Physical Institue of Russian Academy of
Sciences. 119991, Moscow, Russia}
\author{I. G. Lang, L. I. Korovin}
\address{A. F. Ioffe Physical-Technical Institute, Russian
Academy of Sciences, 194021 St. Petersburg, Russia}
\twocolumn[\hsize\textwidth\columnwidth\hsize\csname
@twocolumnfalse\endcsname
\date{\today}
\maketitle\widetext
\begin{abstract}
\begin{center}
\parbox{6in}
{ The average densities of currents and charges, induced by a
weak electromagnetic field in spatially inhomogeneous systems, are
calculated at finite temperatures. The Kubo formula for a
conductivity tensor is generalized for spatially inhomogeneous
systems and fields. The contributions containing electric fields
and derivative from fields on coordinates are allocated. The
Semiconductor quantum wells, wires and dots may be considered as
spatially inhomogeneous systems.}
\end{center}
\end{abstract}

] \narrowtext

\section{Introduction}
In connection with an increased interest to experimental and
theoretical study of low dimensional semiconductor objects -
quantum wells, wires and dots - a construction of the fundamental
theory of interaction of electromagnetic fields with spatially
inhomogeneous systems becomes actual.

R. J. Kubo [1] obtained  the formula for the conductivity tensor
$ \sigma _ {\alpha \beta} (\omega) $, applicable in a case of
spatially homogeneous systems and electric fields $ {\bf E} (t) $
independent of spatial coordinates. This formula takes into
account exactly the interaction of current carriers  with the
medium. Consequently, it is a powerful tool for the solution of
concrete conductivity problems in solids.

 In present work we generalize the Kubo formula on
a case of spatially inhomogeneous systems and  fields. Previously
we calculate average densities of currents and charges induced by
electromagnetic field.

  Kubo [1] used the interaction
operator of current carriers with the electric field as
\begin{equation}
\label {1} U_K =-\sum_ie_i{\bf r}_i{\bf E}(t),
\end{equation}
where $ e_i $ and $ {\bf r}_i $ are the charge and radius -
vector of $ i $ -th particle, respectively, $ {\bf E} (t) $ is the
time-dependent, but spatially homogeneous electric field.
However, it follows from the Maxwell equations that the
time-dependent electric field necessarily depends and on
coordinates, so the use of Eq. (1) is always a certain
approximation, if the field $ {\bf E} $ depends on $ t $. In
spatially inhomogeneous systems the dependence of fields on
coordinates can be essential.

Our task consists in taking into account of heterogeneity of
systems and in obtaining of additional terms in the Kubo formula
 containing derivatives from electric field on coordinates.

The operator of interaction of an electromagnetic field with a
system of charged particles is expressed through the vector $ {\bf
A}({\bf r}, t) $ and scalar $ \varphi({\bf r}, t) $ potentials
(see, for example, [2], page 68), but it is not expressed through
electric $ {\bf E}({\bf r}, t) $ and magnetic $ {\bf H}({\bf r},
t) $ fields, except of individual cases as, for example, a
constant electric field. Accordingly and operators of densities of
 currents $ {\bf j}_1 ({\bf r}, t) $ and charges $
\rho_1({\bf r}, t) $ in linear approximation on vector and scalar
potentials are not expressed through $ {\bf E} ({\bf r}, t) $ and
$ {\bf H} ({\bf r}, t) $. However, average values $ \langle{\bf
j}_1({\bf r}, t)\rangle $ and $ \langle{\bf j}_1({\bf r}, t)
\rangle $ should be expressed through fields, as the observable
values. At finite temperatures $ T $ an average value is defined
as [1,3,4]
\begin{equation}
\label {2} \langle\ldots\rangle=\frac {Sp \{\exp (-\beta {\cal
H})... \}} {Sp \{\exp (-\beta {\cal H}) \}}, \qquad \beta =
\frac{1}{kT},
\end{equation}
$ {\cal H} $ is the Hamiltonian without an interaction of
particles with a weak electromagnetic field.

 The expressions for average $ \langle 0 | {\bf j}_1
({\bf  r}, t) | 0\rangle $ and $ \langle 0 |\rho_1({\bf r}, t) |
0\rangle $ for $ T = 0 $ (when average $ \langle\ldots\rangle $
passes in average $ \langle 0 |\ldots | 0\rangle $ on the ground
state $ |0\rangle $ are obtained in [5]. The present work is a
direct continuation of [5]. The designations and many results of
[5] will be used below.

The problem of an expression of average $ \langle {\bf j} _ 1
({\bf r}, t) \rangle $ and $ \langle \rho_1({\bf r}, t)\rangle $
through electric and magnetic fields is essential also because if
to express average $ \langle{\bf j}_1({\bf r}, t)\rangle $
through vector and scalar potentials, this average contains the
contribution $ - (e/mc)\langle\rho({\bf r})\rangle{\bf A}({\bf
r}, t)) $, where $ e=e_i, m=m_i $ are the charge and mass of the
particle, $ c $ is the light velocity, $ \rho({\bf r}) $ is the
operator of the charge density in a zero approximation on fields
(see Eq. (9) below). This contribution creates complexities at
solutions of some concrete tasks, for example, about light
reflection and absorption by semiconductor quantum wells. These
complexities may be avoided if to express the average $ \langle
{\bf j}_1({\bf r}, t)\rangle $ and $ \langle\rho_1({\bf r},
t)\rangle $ through electric fields and their derivatives on
coordinates.

The question on what kind of interaction to use - containing the
vector potential or electric field - was discussed earlier in [6]
with reference to a task about  light scattering in bulk crystals.
In [6] apparently for the first time the reception is used of
transition from the operators $ {\bf v}_i $ of particles
velocities to the operators $ {\bf r}_i $ of coordinates
according to a quantum ratio $ {\bf v}_i=(i/\hbar)[{\cal H}, {\bf
r}_i] $, due to which it is possible to pass from expressions
containing vector potential to expressions containing electric
fields. However, as we should solve other tasks about spatially -
inhomogeneous systems, it is necessary again to come back to this
theme.

Considering  finite temperatures we use mathematical receptions
offered in [1,3]. We compare our results with conclusions of [4]
devoted to construction of the quantum theory of a spatial
dispersion electric and magnetic susceptibilities. It is supposed
below that there are no charges and currents on the indefinitely
removed distances  and that the fields $ {\bf E}({\bf r}, t) $
and $ {\bf H}({\bf r}, t) $ are equal $ 0 $ on times $
t\rightarrow -\infty $ what corresponds to adiabatic field
switching.

\section{The system Hamiltonian and operators of current and charge densities.}

Let us consider a system of $ N $ particles with the charge $ e $
and the mass  $ m $ in any arbitrary weak electromagnetic field,
characterized by intensities  $ {\bf E}({\bf r}, t) $ and $ {\bf
H}({\bf r}, t) $. Let us introduce vector $ {\bf A} ({\bf r}, t) $
and scalar $ \varphi({\bf r}, t) $ potentials so that
\begin{eqnarray}
\label {3} {\bf E}({\bf r}, t)&=&-\frac{1}{c}\frac{\partial{\bf
A}({\bf r}, t)}{\partial t}-grad\,\varphi ({\bf r}, t),\nonumber\\
{\bf H}({\bf r}, t)&=&rot {\bf A}({\bf r}, t).
\end{eqnarray}
The fields are assumed classical, the gauge of potentials $ {\bf
A} $ and $ \varphi $ is any. For completeness of the task we shall
admit, that the system of particles may be placed in the constant
magnetic field $ {\bf{\cal H}}$ which may be strong. The vector
potential $ {\bf {\cal A}}({\bf r}) $ corresponds to this field,
so that $ {\bf H} = rot {\bf {\cal A}}({\bf r}) $. The total
Hamiltonian $ {\cal H}_{total} $ is as follows
\begin{eqnarray}
\label{4} {\cal H}_{total}&=&\frac{1}{2m}\sum_i({\bf
P}_i-\frac{e}{c}{\bf{\cal A}}({\bf r}_i)- \frac{e}{c}{\bf A}({\bf
r}_i,t))^2\nonumber\\
&+&V({\bf r}_1,\ldots{\bf r}_N)+ e\sum_i\varphi({\bf r}_i,t),
\end{eqnarray}
where $ {\bf P}_i=-i\hbar(\partial/\partial {\bf r}_i) $ is the
generalized momentum operator, $ V({\bf r}_1\ldots{\bf r}_N) $ is
the potential energy, including interaction between particles and
an external potential. In Eq. (4) it is necessary to take into
account non-commutativity of $ {\bf P}_i $ and $ {\bf {\cal A}}
({\bf r}_i),~ {\bf A}({\bf r}_i, t) $. Let us allocate in Eq. (4)
the energy $ U $ of the interaction of particles with
electromagnetic field, including interaction with a strong
magnetic field in the basic Hamiltonian $ {\cal H} $
\begin{equation}
\label {5} {\cal H}_{total}={\cal H} + U,
\end{equation}
where the designations are introduced
\begin{equation}
\label {6} {\cal H} = \frac {1} {2m} \sum _ i {\bf p} _ i ^ 2 + V
({\bf r} _ 1\ldots {\bf r} _ N), ~~ {\bf p} _ i = {\bf P}_
i-\frac {e} {c}{\bf {\cal A}} ({\bf r}_i),
\end{equation}
\begin{eqnarray}
\label{7} U&=&U_1+U_2,\nonumber\\
U_1&=&-\frac{1}{c}\int d^3r{\bf j}({\bf r}){\bf A}({\bf r},t)+
\int d^3r\rho ({\bf r})\varphi({\bf
r},t),\nonumber\\
 U_2&=&\frac{e}{2mc}\int d^3r\rho ({\bf r}){\bf A}^2({\bf
r},t)
\end{eqnarray}
The operators of densities of currents and charges are also
introduced
$$ {\bf j}({\bf r})=\sum_i{\bf j}_i({\bf r}), $$$$
{\bf j}_i({\bf r})={e\over 2}[\delta({\bf r}- {\bf r}_i){\bf v}_
i +{\bf v}_i\delta({\bf r}-{\bf r}_i)], ~~ {\bf v}_i={\bf p}_
i/m, $$
$$\rho({\bf r})=\sum_i\rho_i({\bf r}), \qquad \rho_i({\bf r})=
e\delta ({\bf r} - {\bf r} _ i). $$ The operators $ {\bf j} _ i
({\bf r})$ and $ \rho_i ({\bf r}) $ are connected by the
continuity equation
\begin{equation}
\label{8} div \, {\bf j}_i ({\bf r})+\dot {\rho}_i({\bf r}) = 0,
\qquad \dot {\rho}_i ({\bf r})={i\over\hbar}[{\cal H}, \rho ({\bf
r})],
\end{equation}
which will be used below. The operator $ U_2 $ is out of the
frameworks of linear approximation on fields and is omitted below.

In the Heisenberg representation the additives to the operators of
densities of  currents and charges linear on potentials $ {\bf A}
({\bf r}, t) $ and $ \varphi({\bf r}, t) $ are equal
\begin{eqnarray}
\label{9} j_{1\,\alpha}({\bf r},t)=-\frac{e}{mc}\rho({\bf
r},t)A_\alpha({\bf r},t)\nonumber\\
+\frac{i}{\hbar}\int_{-\infty}^t\,dt^\prime[U_1(t^\prime),
j_\alpha({\bf r},t)],\nonumber\\
 \rho_1({\bf r},t)=\frac{i}{\hbar}
\int_{-\infty}^t\,dt^\prime[U_1(t^\prime),\rho({\bf r},t)],
\end{eqnarray}
where the subscript 1 means a linear approximation on fields, $
\rho ({\bf r}, t), ~~ j ({\bf r}, t) $ and $ U _ 1 (t) $ are
operators in the interaction representation, for example,
$$\rho({\bf r},t)=
e^{i{\cal H}t/\hbar}\rho({\bf r})e^{-i{\cal H}t/\hbar},$$
 $ [F, Q] = FQ-QF $ is the commutator of the operators $
F $ and $ Q $. By substituting  Eq. (7) into Eq. (9) we obtain
for $ U_1 $
\begin{eqnarray}
\label{10} &j&_{1\,\alpha}({\bf r},t)=-\frac{e}{mc}\rho({\bf
r},t)A_\alpha({\bf r},t)\nonumber\\&+& \frac{i}{\hbar
c}\int\,d^3r^\prime\int_{-\infty}^t\,dt^\prime
[j_{1\,\alpha}({\bf r},t),j_\beta({\bf r}^\prime,t^\prime)]
A_\beta({\bf r}^\prime,t^\prime)\nonumber\\ &-&\frac{i}{\hbar}
\int\,d^3r^\prime\int_{-\infty}^t\,dt^\prime [j_{\,\alpha}({\bf
r},t),\rho({\bf r}^\prime,t^\prime)] \varphi({\bf
r}^\prime,t^\prime),
\end{eqnarray}
\begin{eqnarray}
\label{11} &\rho&_1({\bf r},t)=\frac{i}{\hbar
c}\int\,d^3r^\prime\int_{-\infty}^t\,dt^\prime [\rho({\bf
r},t),j_\beta({\bf r}^\prime,t^\prime)] A_\beta({\bf
r}^\prime,t^\prime)\nonumber\\&-&
 \frac{i}{\hbar
c}\int\,d^3r^\prime\int_{-\infty}^t\,dt^\prime [\rho({\bf
r},t),\rho({\bf r}^\prime,t^\prime)] \varphi({\bf
r}^\prime,t^\prime).
\end{eqnarray}

\section{Average values of induced densities of currents and charges.}

 The case $ T=0 $ is considered in
 [5].  And operators of Eqs. (10) and (11)
 are averaged on the ground state of system. It is shown in [7]
(page 84)  that at averaging it is necessary to use wave
functions $ | 0\rangle $ of the ground state without taking into
account the interaction $ U $. Transforming expressions for the
averaged values $ \langle 0 | j_{1\alpha}({\bf r}, t) | 0\rangle
$ and $ \langle 0 |\rho_1({\bf r}, t)|0\rangle $ from [5] we
obtain the results
\begin{eqnarray}
\label{12} \langle 0|j_{1\alpha}({\bf r},t)|0\rangle&=&\langle
0|j_{1\alpha}({\bf r},t)|0\rangle_E+\langle 0|j_{1\alpha}({\bf
r},t)|0\rangle _{\partial E/\partial r},\nonumber\\
 \langle 0|\rho_1({\bf r},t)|0\rangle&=&\langle
0|\rho_1({\bf r},t)|0\rangle_E+\langle 0|\rho_1({\bf
r},t)|0\rangle_{\partial E/\partial r},
\end{eqnarray}
where subscripts $ E $ and $ \partial E/\partial r $ designate
the contributions, containing an electric field and derivative
from field on coordinate, respectively. Passing to consideration
of systems at finite temperatures, we replace in Eq. (12)
averaging $ \langle 0 |\ldots | 0\rangle $ on the ground state by
averaging $ \langle\ldots\rangle $, defined in Eq. (2). Legality
of such replacement will be proved below at comparison with
results of the authors [1,3,4]. Thus, in Eq. (12) we replace
 averaging $ \langle 0 |\ldots | 0\rangle $ by $ \langle\ldots
\rangle $. According to [5] we have
\begin{eqnarray} \label{12}
\label{13} \langle j_{1\alpha}({\bf r},t)\rangle_E
&=&\frac{i}{\hbar}
\int\,d^3r^\prime\int_{-\infty}^t\,dt^\prime\langle
[j_\alpha({\bf r},t), d_\beta({\bf
r}^\prime,t^\prime)]\rangle\nonumber\\
&\times& E_\beta({\bf r}^\prime,t^\prime),
\end{eqnarray}
\begin{eqnarray}
\label{14} \langle \rho_1({\bf r},t)\rangle_E& =&\frac{i}{\hbar}
\int\,d^3r^\prime\int_{-\infty}^tdt^\prime\langle [\rho({\bf
r},t), d_\beta({\bf r}^\prime,t^\prime)]\rangle\nonumber\\
&\times& E_\beta({\bf r}^\prime,t^\prime),
\end{eqnarray}
\begin{eqnarray}
\label{15}\langle j_{1\alpha}({\bf r},t)\rangle_{\partial
E/\partial r}=\frac{e}{mc}\langle d_\beta({\bf r})\rangle
{\partial a_\beta({\bf r},t)\over\partial r_\alpha}\nonumber\\-
 \frac{i}{\hbar
c}\int\,d^3r^\prime\int_{-\infty}^tdt^\prime \langle
[j_\alpha({\bf r},t), Y_{\beta\gamma}({\bf
r}^\prime,t^\prime)]\rangle\nonumber\\
\times {\partial a_\beta({\bf r}^\prime,t^\prime)\over\partial
r_\gamma^\prime},
\end{eqnarray}
\begin{eqnarray}
\label{16} \langle \rho_1({\bf r},t)\rangle_{\partial
E\over\partial r}&=&-\frac{i}{\hbar
c}\int\,d^3r^\prime\int_{-\infty}^tdt^\prime
\nonumber\\
&\times&\langle [\rho({\bf r},t), Y_{\beta\gamma}({\bf
r}^\prime,t^\prime)]\rangle{\partial a_\beta({\bf
r}^\prime,t^\prime)\over\partial r_\gamma^\prime},
\end{eqnarray}
where the designations are introduced
\begin{equation}
\label {17} {\bf d} ({\bf r})={\bf r}\rho ({\bf r}), \qquad Y _
{\beta\gamma} ({\bf r})=r _ \beta \, j _ \gamma ({\bf r}),
\end{equation}
\begin{equation}
\label {18} {\bf a} ({\bf r}, t) \, = \, -c\int _ {-\infty} ^ tdt
^ \prime \, { \bf E} ({\bf r}, t ^ \prime).
\end{equation}
Let us transform the obtained expressions so that the transition
to the Kubo formula  for spatially homogeneous systems and
electric field independent on coordinates would be seen clearly.
We use the ratio [1,3]
\begin{equation}
\label {19} \frac {i}{\hbar} \langle [ F (t), Q (t^
\prime)]\rangle= \int _ 0 ^ \beta \, d\lambda \langle\frac {dQ (t
^\prime)} {dt^\prime}F(t + i\hbar\lambda)\rangle,
\end{equation}
true for any pair operators $ F $ and $ Q $. Using Eq. (19) we
obtain  from Eq. (13)
\begin{eqnarray}
\label{20} &\langle&j_{1\alpha}({\bf r},t)\rangle_E=
\int\,d^3r^\prime\int_{-\infty}^tdt^\prime\int_0^\beta\,d\lambda\nonumber\\
&\times&\langle {\partial d_\beta({\bf
r}^\prime,t^\prime)\over\partial t^\prime} j_\alpha({\bf
r},t+i\hbar\lambda)\rangle E_\beta({\bf r}^\prime,t^\prime).
\end{eqnarray}
It is possible to show that
\begin{equation}
\label {21} {\partial d_\beta ({\bf r}, t)\over\partial t}=-r_
\beta{\partial j_\gamma({\bf r}, t)\over\partial r_\gamma}.
\end{equation}
Substituting Eq. (21) in Eq. (20) and integrating on $ {\bf r}^
\prime $ in parts we obtain
\begin{eqnarray}
\label{22}\langle j_{1\alpha}({\bf r},t)\rangle_E&=&
\int\,d^3r^\prime\int_{-\infty}^tdt^\prime \int_0^\beta\,d\lambda
\nonumber\\
&\times&\langle j_\beta({\bf r}^\prime,t^\prime) j_\alpha({\bf
r},t+i\hbar\lambda)\rangle E_\beta({\bf
r}^\prime,t^\prime)\nonumber\\&+&
 \int\,d^3r^\prime\int_{-\infty}^tdt^\prime
\int_0^\beta\,d\lambda \nonumber\\
&\times& \langle Y_{\beta\gamma}({\bf r}^\prime,t^\prime)
j_\alpha({\bf r},t+i\hbar\lambda)\rangle{\partial E_\beta({\bf
r}^\prime,t^\prime)\over\partial r_\gamma^\prime}.
\end{eqnarray}

In agreement with Eq.(15) the expression for $ \langle j_
{1\alpha}({\bf , r}, t)\rangle_{\partial E/\partial r} $ consists
of two parts. We do not transform first of them, and in the second
we integrate on $ t^\prime $ in parts and afterwards use Eq. (19).
It results in
\begin{eqnarray}
\label{23} \langle j_{1\alpha}({\bf r},t)\rangle_{\partial
E/\partial r}={e\over mc}\langle d_\beta ({\bf r})\rangle
{\partial a_\beta({\bf r},t)\over\partial r_\alpha}\nonumber\\-
{1\over c}\int d^3r^\prime \int_0^\beta\,d\lambda \langle
Y_{\beta\gamma}({\bf r}^\prime) j_\alpha({\bf
r},i\hbar\lambda)\rangle {\partial a_\beta({\bf
r}^\prime,t)\over\partial r_\gamma^\prime}\nonumber\\-
 \int\,d^3r^\prime\int_{-\infty}^tdt^\prime
\int_0^\beta\,d\lambda \langle Y_{\beta\gamma}({\bf
r}^\prime,t^\prime) j_\alpha({\bf
r},t+i\hbar\lambda)\rangle\nonumber\\
\times {\partial E_\beta({\bf r}^\prime,t^\prime)\over\partial
r_\gamma^\prime}.
\end{eqnarray}
Summing Eqs. (22) and (23) we see that last terms in the RHSs of
both formulas are reduced. Total expression we break on two parts
\begin{equation}
\label{24} \langle j_{1\alpha}({\bf r},t)\rangle=\langle
j_{1\alpha}({\bf r},t)\rangle^{(1)}+\langle j_{1\alpha}({\bf
r},t)\rangle^{(2)}
\end{equation}
so, that the first part contains an electric field, and the second
contains a derivative from  field on coordinates, i. e.
\begin{eqnarray}
\label{25} \langle j_{1\alpha}({\bf r},t)\rangle^{(1)}=
\int\,d^3r^\prime\int_{-\infty}^tdt^\prime
\int_0^\beta\,d\lambda\nonumber\\
\times \langle j_\beta({\bf r}^\prime,t^\prime) j_\alpha({\bf
r},t+i\hbar\lambda)\rangle E_\beta({\bf r}^\prime,t^\prime),
\end{eqnarray}
\begin{eqnarray}
\label{26} \langle j_{1\alpha}({\bf r},t)\rangle^{(2)}={e\over
mc}\langle d_\beta ({\bf r})\rangle\, {\partial a_\beta({\bf
r},t)\over\partial r_\alpha}\nonumber\\-
 {1\over c}\int d^3r^\prime \int_0^\beta\,d\lambda \langle
Y_{\beta\gamma}({\bf r}^\prime) j_\alpha({\bf
r},i\hbar\lambda)\rangle {\partial a_\beta({\bf
r}^\prime,t)\over\partial r_\gamma^\prime}.
\end{eqnarray}
It is obvious, that the splitting Eq. (24) does not coincide with
splitting Eq. (12) convenient only at $ T=0 $.

 By similar way we obtain from Eqs. (14) and (16)
\begin{equation}
\label{27} \langle \rho_1({\bf r},t)\rangle=\langle \rho_1({\bf
r},t)\rangle^{(1)}+\langle\rho_1({\bf r},t)\rangle^{(2)},
\end{equation}
\begin{eqnarray}
\label{28} \langle \rho_1({\bf r},t)\rangle^{(1)}=
\int\,d^3r^\prime\int_{-\infty}^tdt^\prime
\int_0^\beta\,d\lambda\nonumber\\
\times \langle j_\beta({\bf r}^\prime,t^\prime)\rho ({\bf
r},t+i\hbar\lambda)\rangle E_\beta({\bf r}^\prime,t^\prime),
\end{eqnarray}
\begin{eqnarray}
\label{29} \langle \rho_1({\bf r},t)\rangle^{(2)}=-{1\over c}\int
d^3r^\prime \int_0^\beta\,d\lambda\nonumber\\
\times \langle Y_{\beta\gamma}({\bf r}^\prime) \rho({\bf
r},i\hbar\lambda)\rangle {\partial a_\beta({\bf
r}^\prime,t)\over\partial r_\gamma^\prime}.
\end{eqnarray}
By obtaining Eqs. (24) - (29) we have reached our main goal: We
have allocated the basic contributions with subscript (1) in
average values of induced densities of currents and charges and
also have shown that the additional contributions with subscript
(2) contain derivatives from, electric field on coordinates. The
sense of splitting on basic and additional contributions is that
at solution of any tasks, in which the field $ {\bf E}({\bf r},
t) $ is spatially inhomogeneous, it is possible to estimate
magnitudes of additional contributions and to determine: It is
necessary to take them  into account or it is possible to reject
them. As well as in the case $ T=0 $, obtained expressions
contain  the operators $ {\bf r}_i $ of particle coordinates
unlike initial Eqs. (10) and (11) which do not contain these
operators.

\section{Comparison with results of [4].}

In [3,4] the expressions for values $ \langle{\bf j}_1({\bf
 r}, t)\rangle $ are received at finite temperatures, but these
expressions differ from deduced by us and above mentioned.
Parallel we make similar calculations of the value $ \langle\rho
_1({\bf r}, t)\rangle $, which  was not considered in [1,3,4],
but in the Maxwell equations it acts on the equal rights with
average density of a current. Let us notice that in [1,3] it was
considered homogeneous medium, in [4] - inhomogeneous. Solving the
equation for the density matrix authors of [3] and [4] come to
the formula, which may be obtained from Eq. (10), if in the RHS
and in the LHS to realize averaging $ \langle\ldots\rangle $
determined in Eq. (2). It is obvious, that similar expression for
$ \langle\rho_1 ({\bf r}, t)\rangle $ may be obtained by averaging
 both sides of Eq. (11).

Further the authors of [3] and [4] transform the expression for
the average induced current density  in such a manner that the
electric field $ {\bf E}({\bf r}, t) $ appears in it and it
contains the vector potential $ {\bf A}({\bf r}, t) $
simultaneously. By doing similar procedure in initial expression
for the average induced charge density we obtain the result of a
kind of Eq. (27), in which the contribution $ \langle\rho_1({\bf
r}, t)\rangle^{(1)} $ is determined by Eq. (28) and the
contribution $ \langle\rho_1({\bf r}, t)\rangle^{(2)} $ is equal
\begin{eqnarray}
\label{30} \langle \rho_1({\bf r},t)\rangle^{(2)}={1\over c}\int
d^3r^\prime \int_0^\beta\,d\lambda \langle j_\beta({\bf r}^\prime)
\rho({\bf r},i\hbar\lambda)\rangle\nonumber\\
\times A_\beta({\bf r}^\prime,t).
\end{eqnarray}
For the contribution $ \langle j_{1\alpha}({\bf r}, t) \rangle^
{(1)} $ from the RHS of Eq. (24) in [3,4] the result of Eq. (25)
is obtained, as well as the expression
\begin{eqnarray}
\label{31} \langle j_{1\alpha}({\bf r},t)\rangle^{(2)}=-{e\over
mc}\langle \rho ({\bf r})\rangle A_\alpha({\bf r},t)\nonumber\\+
 {1\over c}\int d^3r^\prime \int_0^\beta\,d\lambda \langle
j_\beta({\bf r}^\prime) j_\alpha({\bf
r},i\hbar\lambda)\rangle\,A_\beta({\bf r}^\prime,t).
\end{eqnarray}

The authors of [4] went further. With the help of Eq. (31) they
have expressed the derivative from  $ \langle j_{1\alpha}({\bf
r}, t)\rangle^{(2)} $ through an electric field. Integrating this
derivative on time we obtain
\begin{eqnarray}
\label{32} \langle j_{1\alpha}({\bf r},t)\rangle^{(2)}=-{e\over
mc}\langle \rho ({\bf r})\rangle\,a_\alpha({\bf r},t)\nonumber\\+
 {1\over c}\int d^3r^\prime \int_0^\beta\,d\lambda \langle
j_\beta({\bf r}^\prime) j_\alpha({\bf
r},i\hbar\lambda)\rangle\,a_\beta({\bf r}^\prime,t)
\end{eqnarray}
and analogously
\begin{eqnarray}
\label{33} \langle \rho_1({\bf r},t)\rangle^{(2)}={1\over c}\int
d^3r^\prime \int_0^\beta\,d\lambda \langle j_\beta({\bf r}^\prime)
\rho({\bf r},i\hbar\lambda)\rangle\nonumber\\
\times a_\beta({\bf r}^\prime,t).
\end{eqnarray}
Thus, the formulas for the additional contributions of two kinds
are obtained: Eqs. (26) and (29), containing only derivatives of
electric fields on coordinates, and Eqs. (32) and (33), containing
the electric field itself.

Let us show how  it is possible to pass from  Eq. (26) to Eq.
(32), and from Eq. (29) to Eq. (33). In last term from the RHS of
Eq. (26) we integrate on $ r_\gamma^\prime $ in parts. Then we use
the equality
\begin{eqnarray}
\label{34} {dY_{\beta\gamma}({\bf r})\over
dr_\gamma}=j_\beta({\bf r})+ r_\beta(\partial j_\gamma({\bf
r})/\partial r_\gamma)\nonumber\\=j_\beta({\bf
r})-r_\beta\dot{\rho}({\bf r}).
\end{eqnarray}

Then from Eq. (26) we have
\begin{eqnarray}
\label{35} \langle j_{1\alpha}({\bf r},t)\rangle^{(2)}={e\over
mc}\langle d_\beta ({\bf r})\rangle\, {\partial a_\beta({\bf
r},t)\over\partial r_\alpha}\nonumber\\+ {1\over
c}\int\,d^3r^\prime \int_0^\beta\,d\lambda \langle j_\beta({\bf
r}^\prime) j_\alpha({\bf r},i\hbar\lambda)\rangle a_\beta({\bf
r}^\prime,t)\nonumber\\-
 {1\over c}\int\,d^3r^\prime r_\beta^\prime
\int_0^\beta\,d\lambda \langle \dot{\rho}({\bf r}^\prime)
j_\alpha({\bf r},i\hbar\lambda)\rangle a_\beta({\bf r}^\prime,t).
\end{eqnarray}
In the last term we use Eq. (19) and integrate on $ {\bf r}^\prime
$. We obtain that this last term is equal
\begin{eqnarray}
\label{36} -\frac{ie}{\hbar c}\langle [j_\alpha({\bf
r}),\sum_ir_{i\,\beta} a_\beta ({\bf r}_i,t)] \rangle
&=&-\frac{e}{mc}[\langle \rho ({\bf r}) \rangle a_\alpha({\bf
r},t)\nonumber\\&+&\langle d_\beta({\bf r})\rangle {\partial
a_\beta ({\bf r},t)\over\partial r_\alpha}].
\end{eqnarray}
By substituting it in Eq. (35) we come to result of Eq. (32), as
it was required to prove. We transform similarly Eq. (29) for $
\langle\rho_1({\bf r}, t)\rangle^{(2)} $. The difference consists
only that instead of Eq. (36) we have
$$-\frac{ie}{\hbar c}\langle [\rho({\bf r}),
\sum_ir_{i\beta}a_\beta ({\bf r}_i,t)]\rangle=0 $$
 and from Eq. (29) we pass to Eq. (33).

Thus, comparing our results with results of [3] and [4], we have
proved applicability of Eqs. (26) and (29) for the additional
contributions in average values of the induced densities of
current and charge containing only derivatives on coordinates
from electric fields.

\section{The analysis of the formulas for the additional contribution
to the average density of a current.}

The value $ \langle j_{1\alpha}({\bf r}, t)\rangle^{(2)} $ is
determined in three forms of Eqs. (26), (31) and (32). In [4]
some properties of this value are listed. Let us continue its
research and we shall obtain the fourth expression for $ \langle j
_{1\alpha}({\bf r}, t)\rangle^{(2)} $ through derivatives of
vector potential $ {\bf A}({\bf r}, t) $ on coordinates. In the
second term of Eq. (31) for $ j_\beta({\bf r}^\prime) $ we use
Eq. (34). It results in splitting of the second term  into two
parts: In first of them we integrate on $ r _ \gamma ^ \prime $ in
parts, in the second we use the formula
\begin{equation}
\label{37} {i\over\hbar}\langle [F,Q] \rangle\,=\, \int_0^\beta
d\lambda \langle \dot{Q}\,F(i\hbar\lambda) \rangle,
\end{equation}
which follows from Eq. (19) at $t^\prime=t $. By calculating the
commutator we obtain
\begin{eqnarray}
\label{38} \langle j_{1\alpha}({\bf r},t)\rangle^{(2)}= {e\over
mc}\langle d_\beta ({\bf r})\rangle\, {\partial A_\beta({\bf
r},t)\over\partial r_\alpha}\nonumber\\-
 {1\over c}\int\,d^3r^\prime \int_0^\beta\,d\lambda \langle
Y_{\beta\gamma}({\bf r}^\prime) j_\alpha({\bf
r},i\hbar\lambda)\rangle {\partial A_\beta({\bf
r}^\prime,t)\over\partial r_\alpha^\prime}.
\end{eqnarray}

For reception one more - fifth - expression for the additional
contribution to the density of a current we apply Eq. (34) to the
value $ j _ \alpha ({\bf r}, i\hbar\lambda) $ from Eq. (31), and
also following updating of Eq. (37)
\begin{equation}
\label{39} {i\over\hbar}\langle [F,Q] \rangle=\int_0^\beta
d\lambda \langle F\dot{Q}(i\hbar\lambda)\rangle.
\end{equation}
It results in
\begin{eqnarray}
\label{40}\langle j_{1\alpha}({\bf
r},t)\rangle^{(2)}=-\frac{e}{mc} \frac{\partial}{\partial
r_\beta}\{ \langle d_\alpha ({\bf r})\rangle A_\beta({\bf r},t)
\}\nonumber\\+
 \frac{1}{c}\frac{\partial}{\partial r_\gamma}
\left\{\int d^3r^\prime\int_0^\beta d\lambda\langle j_\beta({\bf
r}^\prime)Y_{\alpha\gamma}({\bf r},i\hbar\lambda)\rangle
A_\beta({\bf r}^\prime,t)\right\}.
\end{eqnarray}
From Eq. (40) it follows, that the integral on all space from $
\langle j_{1\alpha}({\bf r}, t)\rangle^{(2)} $ is equal $ 0 $ [4].

\section{The transition to expressions containing a magnetic field.}

Let us obtain the sixth expression for $ \langle j _ {1\alpha}
({\bf r}, t) \rangle ^ {(2)} $, in which we shall introduce a
magnetic field. Let us break Eq. (38) on two parts
\begin{equation}
\label{41} \langle j_{1\alpha}({\bf r},t)\rangle^{(2)}=\langle
j_{1\alpha}({\bf r},t)\rangle^{(-)}+\langle j_{1\alpha}({\bf
r},t)\rangle^{(+)},
\end{equation}
\begin{eqnarray}
\label{42} \langle j_{1\alpha}({\bf
r},t)\rangle^{(\pm)}&=&\frac{e}{2mc} \langle d_{\alpha\beta}({\bf
r})\rangle\left(\frac{\partial A_\beta({\bf r},t)}{\partial
r_\alpha}\pm\right.\left.\frac{\partial A_\alpha({\bf
r},t)}{\partial r_\beta}\right)\nonumber\\&-&
 \frac{1}{2c}\int\,d^3r^\prime\int_0^\beta\,d\lambda
\langle Y_{\beta\gamma}({\bf r}^\prime) j_\alpha({\bf
r},i\hbar\lambda)\rangle\nonumber\\
&\times& \left(\frac{\partial A_\beta({\bf r}^\prime,t)}{\partial
r_\gamma^\prime}\pm\frac{\partial A_\gamma({\bf
r}^\prime,t)}{\partial r_\beta^\prime}\right).
\end{eqnarray}
Since ${\bf H}=rot{\bf A}$, the value $ \langle j_{1\alpha}({\bf
r}, t)\rangle^{(-)} $ is expressed through the magnetic field
\begin{eqnarray}
\label{43} \langle j_{1\alpha}({\bf r},t)\rangle^{(-)}=
-\frac{e}{2mc}({\bf H}({\bf r},t)\times{\bf r})_\alpha\, \langle
\rho({\bf r})\rangle\nonumber\\+
 \frac{1}{2c}\int\,d^3r^\prime\int_0^\beta\,d\lambda
({\bf H}({\bf r}^\prime,t)\,\times\,{\bf r^\prime})_\beta \langle
j_\beta({\bf r}^\prime)j_\alpha({\bf r},i\hbar\lambda)\rangle,
\end{eqnarray}
and  $ \langle j_{1\alpha}({\bf r}, t)\rangle^{(+)} $ may be
expressed through the second derivatives from the vector potential
on  coordinates
\begin{eqnarray}
\label{44} \langle j_{1\alpha}({\bf
r},t)\rangle^{(+)}=-\frac{e}{2mc} \langle \rho({\bf r})\rangle
r_\beta r_\gamma\frac{\partial^2 A_\beta}{\partial
r_\alpha\partial r_\gamma}\nonumber\\+
 \frac{1}{2c} \int\,d^3r^\prime r_\beta^\prime
r_\delta^\prime \int_0^\beta\,d\lambda \langle j_\gamma({\bf
r}^\prime) j_\alpha({\bf r},i\hbar\lambda)\rangle
\frac{\partial^2 A_\beta} {\partial r_\gamma^\prime\partial
r_\delta^\prime}.
\end{eqnarray}
To deduce Eq. (44) from Eq. (42) we have acted as follows: In Eq.
(42) we have used a ratio $ Y_{\beta\gamma} ({\bf r}^\prime)=r_
\beta^\prime j_\gamma({\bf r}^\prime), $ and for $ j_\gamma ({\bf
r}^\prime) $ we have used Eq. (34). Further in the term,
containing $
\partial Y_{\gamma\delta}({\bf r}^\prime)/\partial r_
\delta^\prime, $ we integrated on $ r_\delta^\prime $ in parts
and in the term, containing $ \dot {d}_\gamma ({\bf r}^\prime) $,
we used Eq. (39), integrated on $ {\bf r}^\prime $ and have
calculated the commutator.

It is possible to show that values $ \langle j_{1\alpha}({\bf r},
t) \rangle^{(+)} $ and $ \langle j_{1\alpha}({\bf r}, t)\rangle^
{(-)} $ separately have properties
\begin{equation}
\label{45} div\,\langle {\bf j}_1({\bf r},t)\rangle^{(\pm)}
\,=\,0,\qquad \int\,d^3r \langle j_{1\alpha}({\bf
r},t)\rangle^{(\pm)}=0.
\end{equation}

Let us show that the contribution $ \langle j_{1\alpha} ({\bf r},
t) \rangle ^ {(+)} $ may be expressed through the second
derivatives from an electric field on coordinates. For this
purpose we shall substitute the expression
\begin{equation}
\label{46} {\bf A}({\bf r},t)={\bf a}({\bf
r},t)-c\int_{-\infty}^tdt^\prime{\partial \varphi({\bf
r},t^\prime)\over\partial {\bf r}}
\end{equation}
for the vector potential (following from Eq. (3))  in Eq. (42).
Then in values $ \langle j_{1\alpha}({\bf r}, t)\rangle^{(\pm)} $
it is possible to allocate the contributions from scalar
potential, to which we shall attribute index $ \varphi $, i. e.
\begin{equation}
\label{47} \langle j_{1\alpha}({\bf r},t)\rangle^{(\pm)}=\langle
j_{1\alpha}({\bf r},t)\rangle^{(\pm)}_E+\langle j_{1\alpha}({\bf
r},t)\rangle^{(\pm)}_\varphi.
\end{equation}
At once we obtain that
\begin{equation}
\label{48} \langle j_{1\alpha}({\bf r},t)\rangle^{(-)}_\varphi=0,
\end{equation}
and
\begin{eqnarray}
\label{49} \langle j_{1\alpha}({\bf r},t)\rangle^{(+)}_\varphi=
\int_{-\infty}^tdt^\prime [-{e\over m}\langle d_\beta({\bf}
r)\rangle {\partial^2 \varphi({\bf r},t^\prime)\over\partial
r_\alpha\partial r_\beta}\nonumber\\+
 \int\,d^3r^\prime\int_0^\beta\,d\lambda \langle
Y_{\beta\gamma} ({\bf r}^\prime)j_\alpha({\bf r},i\hbar
\lambda)\rangle {\partial^2 \varphi\over\partial
r^\prime_\beta\partial r^\prime_\gamma}].
\end{eqnarray}
In the second term of Eq. (49) we integrate twice in parts, at
first on variable $ r _ \gamma ^ \prime $, then on variable $ r _
\beta ^ \prime $. Further we use the equation of a continuity Eq.
(8) and Eq. (37). In a result we obtain that the second term from
Eq. (49) is equal to the first with opposite sign and
\begin{equation}
\label {50} \langle j_{1\alpha}({\bf r}, t)\rangle^{(+)}_\varphi
=0.
\end{equation}
Taking into account Eqs. (48) and (50) we obtain from Eqs. (47)
and (42)
\begin{eqnarray}
\label{51}\langle j_{1\alpha}({\bf r},t)\rangle^{(\pm)}&=&\langle
j_{1\alpha}({\bf r},t)\rangle^{(\pm)}_E\nonumber\\
&=&\frac{e}{2mc}\left (\frac{\partial a_\beta({\bf r},t)}{\partial
r_\alpha}\pm\right. \left.\frac{\partial a_\alpha({\bf
r},t)}{\partial r_\beta}
\right)\nonumber\\
&-& \frac{1}{2c}\int d^3r^\prime \int_0^\beta d\lambda \langle
Y_{\beta\gamma}({\bf r}^\prime)j_{1\alpha}({\bf r},i\hbar
\lambda)\rangle\nonumber\\
&\times&\left(\frac{\partial a_\beta({\bf r}^\prime,t)}{\partial
r_\gamma^\prime}\pm\right. \left.\frac{\partial a_\gamma({\bf
r}^\prime,t)}{\partial r_\beta^\prime}\right).
\end{eqnarray}

Taking into account the Maxwell equations $ rot {\bf E} =-(1/c)
(\partial { \bf H}/\partial t) $ and definition Eq. (18) of $
{\bf a} ({\bf r}, t) $ again we come to Eq. (43) for $ \langle j_
\alpha({\bf r}, t)\rangle^{(-)} $, and we obtain also
\begin{eqnarray}
\label{52}\langle j_{1\alpha}({\bf r},t)\rangle^{(+)}=-
\frac{e}{2mc}\langle \rho({\bf r})\rangle r_\beta r\gamma
\frac{\partial^2 a_\beta ({\bf r},t)}{\partial r_\alpha\partial
r_\gamma}\nonumber\\+ \frac{1}{2c}\int
d^3r^\prime\int_0^\beta\,d\lambda r_\beta^\prime
 r_\gamma^\prime\langle j_\gamma({\bf r}^\prime)j_\alpha({\bf
r}, i\hbar \lambda)\rangle \frac{\partial^2 a_\beta ({\bf
r}^\prime,t)}{\partial r_\gamma^\prime
\partial r_\delta^\prime}.
\end{eqnarray}
Taking into account definition Eq. (18) we find that  $ \langle J
_{1\alpha}({\bf r}, t)\rangle^{(+)} $ is expressed through the
second derivatives from an electric field on coordinates. Thus,
the sixth and last formula for the additional contribution $
\langle j_{1\alpha}({\bf r}, t) \rangle^{(2)} $ is determined by
the sum of Eqs. (43) and (52).

\section{The exception of diagonal elements of operators} $ {\bf
r}_i. $

Eqs. (26) and (29) for the additional contributions in average
induced densities of currents and charges, as against Eqs. (32)
and (33), contain the operators $ {\bf r}_i $ of coordinates of
particles. Really, the definitions Eq. (17) may be copied as
\begin{equation}
\label{53} {\bf d}({\bf r})=e\sum_i{\bf r}_i\,\delta ({\bf
r}-{\bf r}_i),
\end{equation}
\begin{equation}
\label{54} Y_{\beta\gamma}={e\over
2}\sum_i\,(r_{i\beta}\,j_{i\gamma}+ j_{i\gamma}r_{i\beta}).
\end{equation}
But average magnitudes $ \langle {\bf j}_1 ({\bf r}, t)\rangle $
and $ \langle\rho_1({\bf r}, t)\rangle $ should not depend on a
point of readout of coordinates $ {\bf r}_i $. It means that Eqs.
(26) and (29) contain only non-diagonal elements of the operators
$ {\bf r}_i $, and the diagonal elements may be excluded. Let us
show that it is so indeed.

Let us transform Eq. (54) so that the operator $ r_{i\beta} $
stood only at the left
\begin{equation}
\label{55} Y_{\beta\gamma}({\bf r})=-{i\hbar\over
2m}\,\delta_{\beta\gamma}\,\rho({\bf r})+
\sum_i\,r_{i\beta}\,j_{i\gamma}({\bf r}).
\end{equation}
Substituting Eqs. (53) and (55) in Eq. (26) we obtain
\begin{eqnarray}
\label{56}\langle j_{1\alpha}({\bf
r},t)\rangle^{(2)}=\frac{i\hbar}{2mc}\int d^3r^\prime\int_0^\beta
d\lambda \langle \rho({\bf
r}^\prime)j_\alpha({\bf r},i\hbar\lambda)\rangle\nonumber\\
\times div\,{\bf a}({\bf r}^\prime,t)+
 \frac{e^2}{mc}\sum_i\langle r_{i\,\beta}\delta ({\bf
r}-{\bf r}_i)\rangle \frac{\partial a_\beta ({\bf r},t)}{\partial
r_\alpha}\nonumber\\
-\frac{1}{c} \int d^3r^\prime\int_0^\beta\,d\lambda\sum_i\,
\langle r_{i\beta}j_{i\gamma}({\bf r}^\prime) j_{1\alpha}({\bf
r},i\hbar\lambda)\rangle\nonumber\\
 \times \frac{\partial a_\beta
({\bf r}^\prime,t)}{\partial r_\gamma^\prime}.
\end{eqnarray}
First two terms in Eq. (56) we shall leave without changes, and
in the last let us split the operator $ {\bf r}_i $ in two parts:
\begin{equation}
\label {57} {\bf r}_i={\bf r}_i^d + {\bf r}_i^{nd},
\end{equation}
where superscripts $ d $ and $ nd $ mean  diagonal and
non-diagonal contributions, respectively. The operator $ {\bf r}_
i^d $ is determined through its matrix elements
\begin{equation}
\label{58} \langle n|{\bf r}_i^d|m\rangle=\langle n|{\bf
r}_i|m\rangle\langle n|m\rangle,
\end{equation}
where $ | n\rangle $ are the eigen-functions of the Hamiltonian $
{\cal H} $. The commutativity property is obvious
$$ [{\cal H}, {\bf r}_i^d]=0. $$
Let us consider the contribution from the operator $ {\bf r}_i^d
$ in the last term in Eq. (56) and designate this contribution as
$ I_\alpha ({\bf r}, t) $. We execute integration on $ r _ \gamma
^\prime $ in parts and use a continuity ratio of Eq. (8). We
obtain
\begin{eqnarray}
\label{59} I_\alpha({\bf r},t)&=&-{1\over c}\int
d^3r^\prime\int_0^\beta d\lambda\sum_i\langle r_{i\beta}^d
\dot{\rho}_i({\bf r}^\prime)j_\alpha({\bf r},i\hbar\lambda)
\rangle\nonumber\\
&\times& a_\beta({\bf r}^\prime,t)).
\end{eqnarray}
As the operator $ {\bf r}_i^d $ commutes  with the Hamiltonian $
{\cal H} $ it is possible to rewrite Eq. (59) as
\begin{eqnarray}
\label{60} I_\alpha({\bf r},t)=-{1\over c}\int
d^3r^\prime\int_0^\beta d\lambda
\nonumber\\
\times\sum_i\langle\dot{\rho}_i({\bf r}^\prime)j_\alpha({\bf
r},i\hbar\lambda) r_{i\beta}^d(i\hbar\lambda)\rangle a_\beta({\bf
r}^\prime,t)).
\end{eqnarray}

Let us substitute in Eq. (60) $ r_{i\beta}^d=r_{i\beta}-r_
{i\beta}^{nd} $. Then the value $ I_\alpha({\bf r}, t) $ will be
split on two parts
\begin{equation}
\label{61} I_\alpha({\bf r},t)=I_\alpha^\prime({\bf r},t)+
I_\alpha^{\prime\prime}({\bf r},t),
\end{equation}
where
\begin{eqnarray}
\label{62} I_\alpha^\prime({\bf r},t)=-{1\over
c}\int\,d^3r^\prime\int_0^\beta\,d\lambda
\nonumber\\
\times\sum_i \langle \dot{\rho}_i({\bf r}^\prime)) j_\alpha({\bf
r},i\hbar\lambda) r_{i\,\beta}(i\hbar\lambda)\rangle a_\beta({\bf
r}^\prime,t).
\end{eqnarray}
\begin{eqnarray}
\label{63} I_\alpha^{\prime\prime}({\bf r},t)={1\over
c}\int\,d^3r^\prime \int_0^\beta\,d\lambda \nonumber\\
\times\sum_i \langle \dot{\rho}_i({\bf r}^\prime) j_\alpha({\bf
r},i\hbar\lambda) r_{i\,\beta}^{nd}(i\hbar\lambda)\rangle
a_\beta({\bf r}^\prime,t).
\end{eqnarray}
Eq. (62) with the help of Eq. (37) turns in
\begin{equation}
\label{64} I_\alpha^\prime({\bf r},t)=-{i\over\hbar c}\int
d^3r^\prime\sum_i \langle [j_\alpha({\bf
r})r_{i\,\beta},\rho({\bf r}^\prime)]\rangle a_\beta({\bf
r}^\prime,t).
\end{equation}
Integrating on $ {\bf r}^\prime $ and calculating the commutator
we obtain the expression
\begin{equation}
\label{65} I_\alpha^\prime({\bf
r},t)\,=\,-\frac{e^2}{mc}\sum_i\langle r_{i\beta} \delta({\bf
r}-{\bf r}_i)\rangle \frac{\partial a_\beta({\bf r},t)} {\partial
r_\alpha},
\end{equation}
which is reduced with the second term in Eq. (56). There remains
the value $ I_\alpha^{\prime\prime}({\bf r}, t), $ which we shall
transform using the continuity equation of Eq. (8) and
integrating on variable $ r_\gamma^\prime $ in parts. It results
in
\begin{eqnarray}
\label{66} I_\alpha^{\prime\prime}({\bf
r},t)=\frac{1}{c}\int\,d^3r^\prime \int_0^\beta\,d\lambda \sum_i
\langle j_{i\gamma}({\bf r}^\prime) j_\alpha({\bf
r},i\hbar\lambda)\nonumber\\
\times r_{i\,\beta}^{nd}(i\hbar\lambda)\rangle \frac{\partial
a_\beta({\bf r}^\prime,t)}{\partial r_\gamma^\prime}.
\end{eqnarray}
Using Eqs. (61), (65) and (66) we obtain the final expression for
the additional contribution to the average induced density of a
current not containing of diagonal matrix elements of the
operator $ {\bf r}_i $:
\begin{eqnarray}
\label{67}\langle j_{1\alpha}({\bf
r},t)\rangle^{(2)}&=&\frac{i\hbar}{2mc}\int
d^3r^\prime\int_0^\beta d\lambda\langle\rho({\bf
r}^\prime)j_\alpha({\bf r},i\hbar\lambda)\rangle\nonumber\\
\times div\,{\bf a}({\bf r}^\prime,t)&+&
 \frac{1}{c}\int d^3r^\prime\int_0^\beta d\lambda
\sum_i \langle j_{i\gamma}({\bf r}^\prime)j_\alpha({\bf
r},i\hbar\lambda)\nonumber\\
&\times&
r_{i\,\beta}^{nd}(i\hbar\lambda)\nonumber\\&-&r_{i\,\beta}^{nd}
j_{i\gamma}({\bf r}^\prime)j_\alpha({\bf r},i\hbar\lambda)\rangle
\frac{\partial a_\beta({\bf r}^\prime,t)}{\partial
r_\gamma^\prime}.
\end{eqnarray}
Similarly we obtain
\begin{eqnarray}
\label{68}\langle \rho({\bf
r},t)\rangle^{(2)}&=&\frac{i\hbar}{2mc}
\int\,d^3r^\prime\int_0^\beta d\lambda\langle\rho({\bf
r}^\prime)\rho({\bf r},i\hbar\lambda)\rangle div\,{\bf a}({\bf
r}^\prime,t)\nonumber\\&+&
 \frac{1}{c}\int d^3r^\prime\int_0^\beta d\lambda
\sum_i\langle j_{i\gamma}({\bf r}^\prime)\rho({\bf
r},i\hbar\lambda)
r_{i\,\beta}^{nd}(i\hbar\lambda)\nonumber\\
&-&r_{i\,\beta}^{nd}
j_{i\gamma}({\bf r}^\prime)\rho({\bf r},i\hbar\lambda)\rangle
\frac{\partial a_\beta({\bf r}^\prime,t)}{\partial
r_\gamma^\prime}.
\end{eqnarray}

\section{The Conductivity Tensor.}

Let us make the Fourier-transformation of an electric field
\begin{equation}
\label{69} E_\alpha({\bf k},\omega)=\int d^3r
\int_{-\infty}^\infty dt E_\alpha({\bf r},t)e^{-i({\bf k}{\bf
r}-\omega t)},
\end{equation}
\begin{equation}
\label{70} E_\alpha({\bf r},t)={1\over(2\pi)^{4}}\int d^3k
\int_0^\infty d\omega E_\alpha({\bf k},\omega)e^{i({\bf k}{\bf
r}-\omega t)}+c.c..
\end{equation}
Average induced density of a current may be written down as
\begin{eqnarray}
\label{71} \langle j_{1\alpha}({\bf r},t)\rangle
={1\over(2\pi)^{4}}\int\,d^3k\int_0^\infty d\omega
\sigma_{\alpha\beta}({\bf k},\omega|{\bf r})\nonumber\\
\times E_\beta({\bf k},\omega)e^{i({\bf k}{\bf r}-\omega t)}+
c.c.,
\end{eqnarray}
where $ \sigma_{\alpha\beta}({\bf k}, \omega | {\bf r}) $ is the
conductivity tensor dependent on spatial coordinates (this
designation is borrowed from [8]).

Using Eqs. (25) and (67) for basic and  additional contributions,
respectively, into the average induced density of a current, we
obtain
\begin{equation}
\label{72} \sigma_{\alpha\beta}({\bf k},\omega|{\bf r})=
\sigma_{\alpha\beta}^{(1)}({\bf k},\omega|{\bf r})+
\sigma_{\alpha\beta}^{(2)}({\bf k},\omega|{\bf r}),
\end{equation}
where
\begin{eqnarray}
\label{73} \sigma_{\alpha\beta}^{(1)}({\bf k},\omega|{\bf r})&=&
\int\,d^3r^\prime\int_0^\infty dt\int_0^\beta d\lambda\langle
j_\beta({\bf r}-{\bf
r}^\prime,-i\hbar\lambda)\nonumber\\
&\times& j_\alpha({\bf r},t)\rangle e^{-i({\bf k}{\bf
r^\prime}-\omega t)},
\end{eqnarray}
\begin{eqnarray}
\label{74} \sigma_{\alpha\beta}^{(2)}({\bf k},\omega|{\bf r})=
\frac{i\hbar k_\beta}{2m\omega}\int d^3r^\prime e^{-i{\bf k}{\bf
r}^\prime} \int_0^\beta d\lambda\nonumber\\
\times\langle\rho({\bf r}-{\bf r}^\prime,-i\hbar\lambda)
j_\alpha({\bf r})\rangle+ \frac{k_\gamma}{\omega}\int d^3r^\prime
e^{-i{\bf k}{\bf
r}^\prime} \int_0^\beta\,d\lambda\nonumber\\
\times\sum_i \langle j_{i\gamma}({\bf r}-{\bf
r}^\prime,-i\hbar\lambda) j_\alpha({\bf r})
r_{i\beta}^{nd}-r_{i\beta}^{nd}(-i\hbar\lambda)\nonumber\\
\times j_{i\gamma}({\bf r}-{\bf r}^\prime,-i\hbar\lambda)
j_\alpha({\bf r})\rangle.
\end{eqnarray}
In the  case $ T=0 $ instead of Eq. (72) we have
\begin{eqnarray}
\label{75} \sigma_{\alpha\beta,0}({\bf k},\omega|{\bf r})=
\sigma_{\alpha\beta}^I({\bf k},\omega|{\bf r})+
\sigma_{\alpha\beta}^{II}({\bf k},\omega|{\bf r})\nonumber\\+
\sigma_{\alpha\beta}^{III}({\bf k},\omega|{\bf r}),
\end{eqnarray}
where
\begin{eqnarray}
\label{76} \sigma_{\alpha\beta}^I({\bf k},\omega|{\bf r})=
\frac{i}{\hbar}\int\,d^3r^\prime \int_0^\infty dt e^{-i({\bf
k}{\bf r}^\prime-\omega t)}\nonumber\\
\times \sum_i\langle 0|j_\alpha({\bf r},t)\rho_i({\bf
r}\nonumber\\-{\bf r}^\prime)
r_{i\beta}^{nd}-r_{i\beta}^{nd}\rho_i({\bf r}-{\bf r}^\prime)
j_\alpha({\bf r},t)|0\rangle,
\end{eqnarray}
\begin{eqnarray}
\label{77} \sigma_{\alpha\beta}^{II}({\bf k},\omega|{\bf r})=
\frac{e^2k_\alpha}{m\omega}\sum_i\langle
0|r_{i\beta}^{nd}\delta({\bf r} -{\bf r}_i)
|0\rangle\nonumber\\
-\frac{ik_\gamma}{\hbar\omega}\int\,d^3r^\prime \int_0^\infty dt
e^{-i({\bf k}{\bf r}^\prime-\omega t)}\nonumber\\\times
 \sum_i\langle 0|j_\alpha({\bf r},t)j_{i\gamma}({\bf
r}-{\bf r}^\prime)
r_{i\beta}^{nd}\nonumber\\-r_{i\beta}^{nd}j_{i\gamma}({\bf r}-{\bf
r}^\prime) j_\alpha({\bf r},t)|0 \rangle,
\end{eqnarray}
\begin{eqnarray}
\label{78} \sigma_{\alpha\beta}^{III}({\bf k},\omega|{\bf r})=
\frac{k_\beta}{2m\omega}\int\,d^3r^\prime \int_0^\infty dt
e^{-i({\bf k}{\bf r}^\prime-\omega t)}\nonumber\\
\times\langle0|[j_\alpha({\bf r},t),\rho({\bf r}-{\bf
r}^\prime)]_+|0\rangle,
\end{eqnarray}
$ [F, Q]_+=FQ + QF $ is the anti-commutator of two operators. It
is possible to pass from Eq. (75) to Eq. (72) if in Eqs. (76) -
(78) to replace averaging $ \langle 0 |\ldots | 0\rangle $ by $
\langle\ldots\rangle $ and to use Eq. (37). \footnote
{\normalsize {In [5] the formula for the conductivity tensor is
given at $ T=0 $, which does not coincide with Eq. (75). This is
because in [5] only diagonal elements $ \langle 0 | {\bf r} _ i |
0\rangle $ are excluded and the operator $ \bar {{\bf r}}_i={\bf
r}_i-\langle 0 | {\bf r}_i | 0\rangle $ is introduced,
distinguished from $ {\bf r}_i^{nd} $.}}

\section{The approximation of a spatially homogeneous electric field.}

In some cases it is possible to neglect the contributions to
average values of the induced densities of currents and charges
containing derivatives from electric field on coordinates, i. e.
to believe
\begin{equation}
\label {79} {\bf E}({\bf r}, t)\simeq{\bf E}(t),
\end{equation}
as it is made, for example, in [1] though, strictly speaking,
spatially homogeneous field $ {\bf E} $ may be only
time-independent.
 In approximation Eq. (79) the Fourier-transformation is
\begin{equation}
\label{80} E_\alpha(\omega)=\int_{-\infty}^\infty dt e^{i\omega
t}E_\alpha(t).
\end{equation}
Then it is possible to write down
\begin{equation}
\label{81} \langle j_{1\alpha}({\bf r},t)\rangle_h={1\over
2\pi}\int_0^\infty d\omega\sigma_{\alpha\beta}(\omega|{\bf
r})E_\beta(\omega)e^{-i\omega t}+ c.c.,
\end{equation}
where the subscript $ h $ means a spatially homogeneous field. It
is easily to see that
\begin{equation}
\label{82} \sigma_{\alpha\beta}(\omega|{\bf r})=
\sigma_{\alpha\beta}({\bf k}=0,\omega|{\bf r}).
\end{equation}
Then with the help of Eqs. (72)-(74) we obtain
\begin{equation}
\label{83} \sigma_{\alpha\beta}(\omega|{\bf r})=\int_0^\infty dt
\int_0^\beta\, d\lambda \langle
J_\beta(-i\hbar\lambda)j_\alpha({\bf r},t)\rangle e^{i\omega t},
\end{equation}
where
\begin{equation}
\label{84} J_\alpha=e\sum_i\,\dot{r}_{i\alpha}
\end{equation}
is the current operator.

Eq. (83) is the generalization of the Kubo formula for the
spatially inhomogeneous medium when the conductivity tensor
depends on $ {\bf r} $.

Further we shall consider a case of a spatially homogeneous medium
 in which any average values can not depend on coordinates $
{\bf r} $. Then the tensor $ \sigma_{\alpha\beta} (\omega | {\bf
r}) $ does not depend on $ {\bf r} $ and we obtain from Eq. (83)
\begin{equation}
\label{85} \sigma_{\alpha\beta}(\omega)={1\over V_0}\int_0^\infty
dt \int_0^\beta d\lambda\langle
J_\beta(-i\hbar\lambda)J_\alpha(t)\rangle e^{i\omega t},
\end{equation}
where $V_0$ is the normalization volume. The obtained formula
coincides with the Kubo result [1] if to replace in the RHS  $
\omega $ on $ -\omega $ and to take into account that in [1]  $
V_ 0 = 1 $.

\section{The case of a constant magnetic field.}

Let us consider a case when an external weak electromagnetic
field is reduced to a constant in space and time-independent
magnetic field $ {\bf H}({\bf r}, t)={\bf H} $ and an electric
field $ {\bf E} ({\bf r}, t) = 0 $. Let us remind that we have
included the vector potential $ {\bf{\cal A}}({\bf r}), $
appropriate to a constant magnetic field $ {\bf H}$, in the basic
Hamiltonian Eq. (6). However, in the present section we believe $
{\bf H}= 0, \, {\bf{\cal A}}({\bf r})=0 $ and the field $ {\bf
H}=const $ is considered so weak that it is possible to be
limited by linear on a field  contributions to the induced
densities of currents and charges. Then, according to Eq. (25),
the basic contribution to induced density of a current  $ \langle
{\bf j}_1 ({\bf r}, t) \rangle ^ {(1)}=0 $, since $ {\bf E} ({\bf
r}, t)=0 $. Let us choose the vector potential as
\begin{equation}
\label{86} {\bf A}({\bf r})={1\over 2}({\bf H}\times{\bf r}).
\end{equation}
Then it is obvious from Eq. (44) that $ \langle{\bf j}_1({\bf r},
t)\rangle^{(+)}=0 $ since it contains second derivatives from $
{\bf A} ({\bf r}) $ on coordinates. There is only contribution $
\langle{\bf j}_1({\bf r}, t)\rangle^{(-)} $ determined in Eq.
(43). Thus, at $ {\bf H} = const $ in linear approximation on the
field we managed to express the density of the induced current
through the magnetic field intensity. Now we shall exclude
diagonal matrix elements of the operators $ {\bf r}_i $ from
expression for $ \langle{\bf j}_1({\bf r}, t)\rangle $ at $ {\bf
H}=const. $ For this purpose let us take advantage of Eq. (67) in
which vector $ {\bf a} ({\bf r}, t) $ can be replaced by vector
potential $ {\bf A} ({\bf r}, t) $, since the initial expression
of Eq. (26) can be replaced by Eq. (38). By substituting Eq. (86)
in Eq. (67) we obtain
\begin{eqnarray}
\label{87} \langle j_{1\alpha}({\bf
r},t)\rangle=-\frac{e^2}{2mc}\sum_i \langle ({\bf H}\times{\bf
r}_i^{nd})_\alpha\delta({\bf r}-{\bf r}_i)\rangle
\nonumber\\-\frac{ie}{2\hbar c}\sum_i\langle j_\alpha({\bf
r})H_\beta({\bf r}_i^{nd}\times{\bf
r}_i^{nd})_\beta\rangle\nonumber\\+
 \frac{e}{2c}\int_0^\beta d\lambda \sum_i \langle
({\bf H}\times {\bf r}_i^{nd})_\beta\, v_{i\beta}j_\alpha({\bf
r},i\hbar\lambda)\rangle.
\end{eqnarray}
Let us notice that the vector $ {\bf r}_i^{nd} \times {\bf r}_i^
{nd}\neq 0 $ because projections $ r_{i\alpha}^{nd} $ with
different subscripts $ \alpha $ do not commute among themselves,
for example,
\begin{equation}
\label{88} ({\bf r}_i^{nd}\times{\bf r}_i^{nd})_z=[{\bf
r}_{ix}^{nd},{\bf r}_{iy}^{nd}].
\end{equation}

\section{Conclusion.}

Let us list the obtained basic results. It is shown that average
values of densities of currents and charges induced by weak
electromagnetic field at finite temperatures and spatially
inhomogeneous systems are expressed through electric fields and
their derivatives on coordinates. The contributions expressed
through electric field were called "basic" and through derivatives
- "additional".

For the "additional" contributions to average values of induced
densities of currents and charges six pairs of various
expressions are obtained. Two of these expressions for the density
of a current coincide with the results of [4]. But the expressions
from [4] contain electric fields or vector potentials, instead of
derivatives from these values on coordinates, that complicates an
estimation of value of the "additional" contributions. Generally
speaking, integrating on $ {\bf r}^\prime $ in parts it is
possible to get rid of derivatives $ \partial E_\beta({\bf r}^
\prime, t)/\partial r_\gamma^\prime $ passing to the formulas
containing only fields instead of derivative from them. However,
the opposite procedure - transition from a field to derivatives -
is not possible always. In the "basic" contributions the fields
are always kept.

The sixth expression obtained in section VI for the "additional"
contribution to average induced density of a current breaks up on
two parts. First of them with an index $ (-) $ is expressed only
through magnetic field $ {\bf H} ({\bf r}, t) $, the second with
an index $ (+) $ - through the second derivative from an electric
field on coordinates. The similar result is obtained for the
average induced density of charge. If the "additional"
contributions are expressed through derivatives from electric
fields, the operators $ {\bf r}_i $ of coordinates of particles
enter in the appropriate formulas necessary. It may seem that
this result is absurd, since coordinate $ {\bf r}_i $ depends on
a point of a beginning of readout. However, it appears that
diagonal matrix elements $ \langle n | {\bf r} _ i | n\rangle $
do not enter into the average induced densities of currents and
charges, that is shown in section VII, and the non-diagonal
elements do not depend on a beginning  point of readout .

In section VIII the "basic" and "additional" contributions are
calculated in the conductivity tensor for spatially inhomogeneous
systems and fields. In approximation, when the electric field is
homogeneous in space, but time-dependent (section IX), the "
basic " contributions are kept only. For this case for the
conductivity tensor, dependent on frequency $ \omega $ and
coordinates $ {\bf r} $, the modified Kubo formula is obtained
which passes in the formula from [1] for spatially homogeneous
systems.

At last, in section X the expression for the average induced
density of a current is obtained in a case when a weak
electromagnetic field is reduced to a constant magnetic field.

It follows from Eq. (74) that the "additional" contributions in
conductivity contain a factor $ k_\gamma/\omega $. If a field $
{\bf E} ({\bf r}, t) $ is the plane wave extending with the light
velocity $ c $ (at a monochromatic irradiation) or the wave
package (at a pulse irradiation), then $ k\simeq\omega/c $ and
additional contributions contain in comparison with basic a small
factor $ v/c $, where $ v $ is the speed of particles in system.
However, this estimation is not correct always in case of
spatially inhomogeneous systems, for example, in semiconductor
quantum wells, wires or dots. It is possible to consider a field $
{\bf E}({\bf r}, t) $ as an external or stimulating field only in
the case of calculating the density of the induced current and
charge in the lowest order on interaction of a field with system
of charged particles. Such approximation is allowable in case of
quantum wells under condition of [9,10] $ \gamma _ r\ll \gamma, $
where $ \gamma_r(\gamma) $ is the radiative (non-radiative)
broadening of electronic excitations.

Otherwise, when $ \gamma_r\gg \gamma, $ it is necessary to take
into account interaction of a field with particles in all orders
of the perturbation theory, and then $ {\bf E} ({\bf r}, t) $ is
the genuine field within the low dimensional object. This field
already cannot be presented as a superposition of plane waves for
which $ k = \omega/c $. For example, genuine field strongly
varies  within  a quantum well along an axis $ z $, perpendicular
to the well plane, if light is directed along an axis $ z $, and
the frequency $ \omega $ is in a resonance with one of discrete
energy levels of the electronic system excitations in a quantum
well [11,12]. Then the values $ kd\simeq 1 $, where $ d $ is the
quantum well width, instead of small factor $ v/c $ appears  a
the factor
$$M\simeq\frac{v}{\omega d}=\frac{v\lambda}{2\pi cd}.$$
If  a wave length  $ \lambda\gg d $ it can appear that the new
factor
 $ M $ is greater than $ v/c $. In concrete cases
one needs to estimate its value. In [11,12] it was supposed that
$ M\ll 1 $ and  the "additional" contributions to average
 induced densities of currents and charges
(the case $ T=0 $ was considered ) were neglected. By substituting
the received expressions for average densities of a currents and
charges in the Maxwell equations, basically it is possible to
determine true fields inside and outside of low-dimension
semiconductor objects. Thus, it is possible to calculate factors
of reflection and absorption of light by these objects (see, for
example, [11,12]).

The authors are grateful M. Harkins for a critical reading of the
manuscript.

\end{document}